# Predictive Analytics Using Smartphone Sensors for Depressive Episodes

Taeheon Jeong, Northwestern University
Diego Klabjan, Industrial Engineering and Management Sciences, Northwestern University
Justin Starren, Chief, Division of Health and Biomedical Informatics, Northwestern University

**Abstract**

The behaviors of patients with depression are usually difficult to predict because the patients demonstrate the symptoms of a depressive episode without a warning at unexpected times. The goal of this research is to build algorithms that detect signals of such unusual moments so that doctors can be proactive in approaching already diagnosed patients before they fall in depression. Each patient is equipped with a smartphone with the capability to track its sensors. We first find the home location of a patient, which is then augmented with other sensor data to identify sleep patterns and select communication patterns. The algorithms require two to three weeks of training data to build standard patterns, which are considered normal behaviors; and then, the methods identify any anomalies in day-to-day data readings of sensors. Four smartphone sensors, including the accelerometer, the gyroscope, the location probe and the communication log probe are used for anomaly detection in sleeping and communication patterns.

## I. Introduction

### A. Motivation & Problem Statement

Depression population, as of 2012, has exceeded 350 million worldwide, according to the World Health Organization (Marcus and Yasamy 2012). Depression is different from simple mood fluctuations or temporary emotional responses because this disease can cause severe consequences to humans. Typical symptoms of depressions can range widely from less severe ones such as difficulty concentrating and fatigue, to more serious indications such as persistent aches, feeling of anxiety and "emptiness." In extreme cases, depression can lead to suicidal attempts and in fact, Centers for Disease Control and Prevention reports depression as one of the leading causes of suicides.

Even more critical issues are behind the fact that depression symptoms do not appear regularly. The patients function normally for the most of the time but become depressed at unexpected moments. Due to the inconsistent nature of the disease, doctors may fail to preempt the symptoms of a forthcoming depressive episode at the right times and struggle to take appropriate measures in a timely basis. For this reason, this research is to develop algorithms that automatically raise an alert to a doctor when his or her patients are going through initial symptoms of a depressive episode.

Increasingly, smart phones are being used for behavioral therapy, including in patients with depression. While self-reported symptoms will continue to be a critical part of monitoring for these patients, self-reported data is always subject to a certain level of bias. Supplementing self-reported data with behavior data that tracks changes in activity will be an important component of comprehensive monitoring programs.

We assume that the patient has already been diagnosed with depression and equipped with a smart phone. The goal is to predictively preempt episodes of depression attacks.

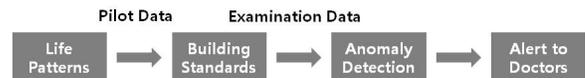

**Figure 1.** Flow diagram of how the abnormal behavior of patients is detected and reported to doctors.

### B. Proposed Approach

Two noticeable symptoms of depressive episodes are the changes in sleep patterns, either insomnia (if a person wakes up too early) or hypersomnia (if a person oversleeps), and the changes in daily activities, especially the loss of interest in social activities. In this research, we have exploited these two apparent changes in life patterns to determine a patient's state.

To do so, three algorithms have been developed. The location probe, also known as GPS, helps locating the patient's home location, which is essential to verifying the sleep patterns under the premise that people mostly sleep at home. The second algorithm utilizes the accelerometer and gyroscope to keep track of the motions of the smartphone. The home location and the device motion records return in-formation on sleep patterns. The third algorithm works with the communication log probe and helps determine user's involvement in social activities.

### C. Contributions and Prior Work

As the smartphones become more widespread, studies using data sets from smartphone sensors are also emerging. Recently, Ginger.io Inc., has introduced a smartphone application that finds the times when a user is in depression. However, there is a clear distinction in that this application focuses only on communication patterns and travel distances; in contrast, our research largely concentrates on finding the users' sleep patterns. No details are known to the authors about the underlying methodologies of Ginger.io.

For the past decades, significant efforts have been invested to find algorithms that detect depression. For instance, there had been several studies on human actions to identify depression symptoms. A research on facial actions and vocal prosody has achieved accuracy of 88% in detecting the symptoms (Cohn et.al. 2009). Jarrold et.al. [2010] conduct research on brain health by using data mining tools to identify clinical depression and have also achieved a high range of accuracy from 73% to 97%. However, the outcomes of these studies are not able to detect the symptoms without interactions between the doctor and patient. In essence, they are reactive in nature as opposed to our study that focuses on being proactive and detection of depressive episodes. Smartphone based analysis can eliminate the needs for such interactions.

The two key patterns that our research focuses on are the sleep and communications patterns. There also have been studies that monitor these patterns. A research on sleep quality has utilized smartphones to detect sleep patterns using noise sensors (Hao et.al. 2013). However, their research mostly focused on sound aspects, such as snoring, unlike sleep start hours and duration as is the case in our research. Communication patterns had also been studied. Similar to our research, Candia's et.al. [2008] research has built an algorithm that detects anomalies in the communication patterns using phone records. Our research is clearly different from theirs in that we focus specifically on depressive episode symptoms and our algorithms are based on the detection of both sleep and communication abnormal behaviors.

There has also been research that shares the exact same purpose as ours (proactively detecting depressive episods) but approaches the problem differently. Doryab's et.al. [2014] study investigates social and sleep behaviors to detect any behavior changes using smartphones but their algorithm depends on correlation analysis whereas, our research uses clustering method and other different statistical tools. Another research done by Mok et.al. [2014] also builds a depressive episode detector. However, Mok's et.al. research examines the relationship between voice characteristics and depression instead of the relationship between sleep, communication and depression as we do.

## II. Sensors

### A. Location Probe

Commonly known as GPS, the location probe monitors the smartphone user's location in latitude and longitude. The rate at which the data is collected can differ depending on the settings of the device.

### B. Accelerometer

Accelerometer measures either the static or dynamic forces the sensor is experiencing in $x$, $y$ and $z$ directions. For instance, when a smartphone is laid down flat, the accelerometer returns a value close to (0,0,9.8), the $z$ value representing the gravity force. On the other hand, when the smartphone moves in the $x$ direction with acceleration of $2 \text{ m/s}^2$, the sensor should show a value close to (2,0,9.8). The actual measurements will never be exactly 2,0 or 9.8 in examples above due to noises.

### C. Gyroscope

Gyroscope is a sensor that uses Earth's gravity to help determine the orientation. Unlike accelerometer, gyroscope measures the rotations of a device in $x$, $y$ and $z$ directions. For instance, when a smartphone is laid down flat, the gyroscope returns a value close to (0,0,0). When the smartphone rotates in the $x$ direction with 0.2 rad/s, the sensor should show (0.2,0,0). Similar to the accelerometer, the actual measurements are approximates due to noises.

### D. Communication Log Probe

The communication log probe in a smartphone mainly monitors phone calls and SMS. Incoming, outgoing and missed calls, and incoming and outgoing SMS are the five essential numbers collected from this log.

## III. Model Training

In this section, we describe how normal behavior patterns can be built from two to three weeks of training data. K-means clustering and confidence intervals are key methodologies in this portion of the research. We assume that each patient is given a smartphone and that during the training period the patient does not have depressive symptoms. This assumption is reasonable since for training a few weeks of data are needed while periods between episodes are usually longer than a month.

## A. Home Location

The location probe typically records the latitude and longitude as often as every second. Figure 2 shows locations of a test subject. High density of points suggests that the smartphone user spent a long period of time at the same location; hence, the location could represent places like home, work or school. In contrast, locations with low density of points would be much less significant.

To take advantage of these density differences, the home location algorithm uses the k-means clustering method. Figure 3 shows a clustering while Figure 4 is the underlying scree plot justifying three clusters in this case because there is a significant drop in the sum of square errors from 3 clusters to 4 clusters. Note that 4 clusters would create a tighter clustering but at the expense of increased complexity. The scree plot shows that the trade-off is achieved for 3 clusters. We still need a method to distinguish the home cluster among all other clusters. The home location algorithm only uses those data points that are recorded during night hours, from midnight to 6am by default. Assuming that the smartphone user lives with a typical life pattern (work or school during the day and home during the night), we can infer that the largest cluster of all will be the home cluster. During the night hours, the home cluster should be self-evident. In atypical cases such as night shift workers, the algorithm is modified to select the data only from the day time. In what follows for ease of exposition we assume night time sleeping. Once the home cluster is chosen, the algorithm simply assigns the center of cluster to be the home latitude and longitude.

In clustering it is critical to choose an appropriate number of clusters to form. For this, we use the scree plot. The detection of the number of clusters is automated by selecting a point on the curve where the slope decreases by a substantial relative value.

The latitude and longitude of the largest cluster chosen from the appropriate number of clusters during the night hours then becomes the home location.

## B. Sleep Pattern

The accelerometer and gyroscope provide very useful information that determines the sleep pattern. Due to the challenges discussed later, the sleep pattern algorithm finds the longest interval during a day when the smartphone does not move and proclaims this as the sleep interval.

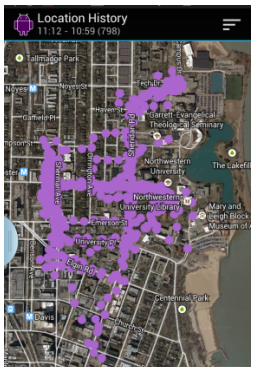

**Figure 2.** A test user's sample location history

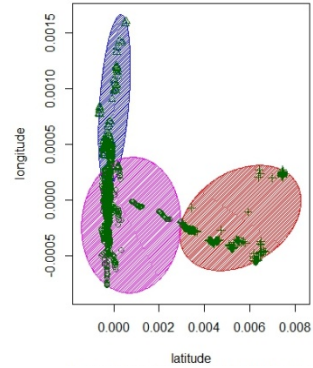

**Figure 3.** The result of k-means clustering with appropriate number of clusters of three

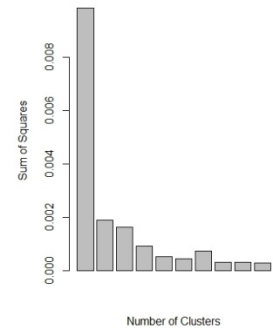

**Figure 4.** A sample scree plot that justifies three clusters.

However, because there could be multiple such long intervals, the longest interval we find may not necessarily match with the sleep interval. For instance, a smartphone user may not be an avid user and he/she may leave the device not moving for a long time throughout the day. In such a case, the longest interval of a smartphone laying down flat does not necessarily imply a sleep period.

Two assumptions allow us to find a more accurate sleep interval. A person sleeps at home and at night (note that the latter assumption is already made by the home location detection algorithm). Given these two assumptions, the algorithm finds the longest interval of smartphone not moving when the device is located at home during night hours. The algorithm is modified to accommodate atypical cases such as night shift workers or people who spend extended periods of time away from home.

As previously mentioned, an accelerometer returns a value close to (0,0,9.8) and a gyroscope returns a value close to (0,0,0) when the smartphone lays flat. Therefore, the longest interval that the data records do not deviate from these values is the sleep interval. Sensors can never be perfect because of the noise; hence, different error tolerances depending on the device must be taken into account.

Once the sleep intervals are determined throughout the weeks of the training data, we perform another k-means clustering to identify sleeping patterns (which for example vary by the day of the week). The sleep interval consists of two components: sleep start hour on the *x* axis and sleep duration on the *y* axis. Similar to the home location algorithm, the scree plot suggests the proper number of clusters. On the other hand, with the sleep intervals, we do not find the largest cluster because the user may have several different sleep patterns depending on the day of the week.

For instance, a person may sleep from midnight for six hours from Sunday to Thursday and sleep from 10 pm for eight hours on weekends. In this case, we do not want to select the largest cluster but keep the two separate clusters because they both represent normal behavior.

Events unrelated to depression may produces instances of abnormal sleep during the two to three weeks of training data collection. Standard patterns built during those weeks should only contain normal behaviors because these patterns are used for later inspection of anomalies. The algorithm therefore adapts an outlier detection algorithm to avoid any abnormal behaviors in the standard pattern by discarding sparse clusters far from dense clusters.

Figure 5 shows three distinctive clusters including a small single data point cluster in the bottom right corner. The two big clusters represent well behaved sleep patterns; in contrast, the small cluster (bottom right corner) is a clear deviation from the two normal patterns. In such case, the outlier algorithm will remove the small cluster and maintain only the two large clusters.

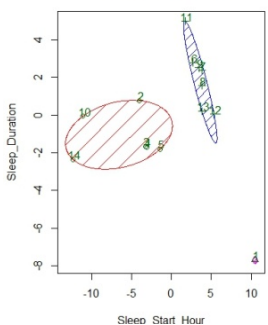

**Figure 5.** A sample cluster plot with an outlier on the bottom right corner.

### C. Communication Patterns

One symptom associated with depressive episodes is the loss of interest in social activities. The level of social activeness is well reflected in the use of smartphones for communication. Hence, we can infer that the engagement with the smartphone would decrease when the user is in an early stage of a depression.

The key components of the communication data are the call and SMS logs. The depressive episode symptoms should not be correlated with the incoming calls or the incoming SMS because a smartphone user does not have any control over them. Therefore, the communication algorithm utilizes numbers collected for outgoing and missed calls and outgoing SMS only.

For this task, a confidence interval calculated from the train data sets the regular pattern for later evaluations of anomalies. The confidence interval assesses whether the user's phone call and SMS usages are within a reasonable range. The algorithm implements the 95% confidence level.

When calculating the confidence interval, the algorithm first calculates the "delta" values for all outgoing calls, missed calls and outgoing SMS. The "delta" values are the daily changes in aggregate values. For instance, 10 outgoing SMS one day and 8 outgoing SMS the next day give a delta of -2. Because we need to find times when social activeness shows dramatic changes, the delta values are more relevant than the raw values directly obtained from the logs. Figure 6 shows a sample graph of the delta values and it demonstrates two dramatic drops in social activities on the second and the third day.

The challenge in calculating the confidence interval is that the delta values may be small and it could be hard to obtain a good interval from small delta values. The phone and SMS usage for most people do not fluctuate too much from day to day and this may result in very small delta values. For this reason, the algorithm sums three consecutive days worth of data and uses these values to build the confidence interval based on deltas. For example, if there have been 5, 3, 4, 2 outgoing calls on a Monday, Tuesday, Wednesday and Thursday, respectively, we first sum values from Monday to Wednesday and also the values from Tuesday to Thursday. These aggregated total values are used to calculate the delta values and eventually the confidence interval.

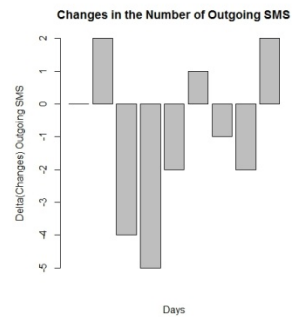

**Figure 6.** An example of delta values of outgoing SMS

One last factor that we have to consider is the change in patterns on weekdays and weekends. For some people, social activities may increase over the weekends or for others, it could be the opposite. Therefore, the algorithm is able to evaluate the weekday and weekend data separately. In such case, the three-day sum would only be applicable to the weekdays but not weekends and there would be two separate confidence intervals for weekdays and weekends.

Figure 7 is an example of the confidence interval obtained from the three-day sum values. The confidence interval obtained from the train data is later used to evaluate the state of the smartphone user, i.e. patient.

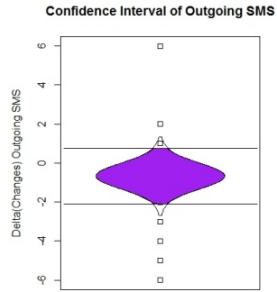

**Figure 7.** A sample 95% confidence interval of the communication delta values, anything below the colored region may signify abnormal behaviors.

## IV. Anomaly Detection on Real Time

In this section, we discuss how anomalies are detected in real-time from model calibration based on the train data. The home location will not be a subject of evaluation because it is only needed to help finding the sleep patterns. Therefore, only the sleep and communication patterns are used.

### A. Sleep Pattern

Once calibrations are completed from the train data, it is enough for the smartphones to collect the accelerometer records and the gyroscope data in real time. Since, we obtained the exact home location from the train data, there is no need for finding it again.

Analogous to calibration, the algorithm finds the longest interval when the phone lays down flat while the device is located at home at night, to figure out when and for how long this patient has been sleeping on the day in question. Since we formed clusters out of the sleep start hour and the sleep duration, the algorithm determines whether the user's sleep pattern on the day in question fits into any one of the clusters from the calibration. If not, it indicates abnormal behavior.

### B. Communication Pattern

Just as in training, we gather in real time the same three items: outgoing calls, missed calls and outgoing SMS. Furthermore, because calibration is based on the delta concept for three consecutive days, we do the same for real-time data. In the case which the algorithm treats weekdays and weekends differently as discussed previously, we would also make a distinction between them with test data as well.

With the calculated confidence interval, the algorithm would treat different components of the data differently. For outgoing calls and SMS, we disregard any value above the upper limit because we are trying to find the times when the patient is socially inactive. Therefore, on a day-to-day basis, the algorithm sums up the values of today and the previous two days to see if the delta value drops below the lower limit of the confidence interval. If so, it is likely to indicate social inactiveness. On the other hand, we also want to detect when the number of missed calls significantly rises. Hence, in this case, we can ignore too few missed calls that are below the confidence interval, but an anomaly occurs when the number of missed calls for the last three days raises over the upper limit of the interval. As a result, the times when the numbers of outgoing calls or SMS are too small or the times when the number of missed calls dramatically rises indicate abnormal behavior. In the case of distinction between weekdays and weekends, the same logic would apply to weekends, except everything will be evaluated based on daily values instead of three day sums.

## V. Study

This section provides a sample study of the sleep and communication pattern evaluations. The system consists of the Android application PurpleRobot (Schueller et.al. 2014 and [1]), which is installed on each smartphone provided to a patient. The application streams the sensor data to a central database in mongoDB. For training, the data over a period of three weeks is downloaded to a relational database (postgres) and the calibration is performed in R statistics.

This study represents data from normal volunteers in an attempt to establish normal baselines to support future research. We discuss sample cases next.

### A. Sleep Pattern

Figure 8 shows a sample result of sleep pattern evaluations. In this specific example, we exhibit six days. The "cluster" column shows the cluster that the new data belongs to. We notice that on the sixth day, the behavior of the test user was clearly a deviation from the norm. The user went to sleep at around 11:30am in the morning and slept for about an hour and half. The cluster assigned to this day was "none," which indicates that this day does not belong to any of the clusters formed.

---

[1] CBITs TECH Website, PurpleRobot. URL: http://tech.cbits.northwestern.edu/purple-robot/

### B. Communication Pattern

Figure 9 shows an evaluation result of weekday communication patterns. The "P/F" column gives information about the state of the smartphone user. Value of 1 indicates that the delta value is within the confidence interval. In contrast, value of 0 suggests anomaly. Similarly, Figure 10 is a sample outcome for evaluations on weekend days, in case the user decides to evaluate weekdays and weekends separately.

|   | Sleep Start Hour | Sleep Duration | Cluster |
|---|---|---|---|
| 1 | 3.73333 | 5.73748 | 2 |
| 2 | 1.53333 | 4.74440 | 2 |
| 3 | 0.88333 | 5.85701 | 2 |
| 4 | 2.71667 | 4.02897 | 2 |
| 5 | 3.76667 | 2.97549 | 2 |
| 6 | 11.41667 | 1.51673 | Outlier |

**Figure 8.** A sample result of the sleep pattern evaluations

## VI. Challenge and Improvements

Although the sleep and communication pattern we find with the algorithms can give us useful information about the smartphone user's state, there exist limitations to this research.

The biggest challenge is that the smartphone is an excellent source of data but not the perfect tool for finding a sleep pattern. For instance, to find the sleep start hour, the algorithm looks for an instance when the phone stopped moving. However, for some people, smartphones are not the very last thing that they interact with before going to sleep. Similarly, smartphone might not be the first thing people look for when they wake up. As a result, the biggest challenge is that we can never accurately figure out the exact sleep start hour and sleep duration from smartphones. This is the reason why other similar studies try to deploy wearable devices rather than smartphones. These facts do not necessarily adverse our study, since instead of detecting sleep patterns, we can interpret our algorithms as detecting anomalies in phone use for extended periods of daily inactivity. This also serves the purpose for depression patients.

There are possible improvements to our methodology. Smartphones nowadays are equipped with more than twenty sensors. Possible options that could improve our methodology include the light sensor or proximity probe. The light probe detects the brightness of the environment in Lux. However, this can also encounter challenges distinguishing between indoor and outdoor darkness. The proximity probe can play a pivotal role because it can directly tell whether the user is engaged with the phone or not. When the user is using the phone the proximity probe should record small distance values.

## VII. Acknowledgements

|   | weekday 1 | weekday 2 | weekday 3 | Outgoing Calls | Delta Calls | Calls P/F | Outgoing SMS | Delta SMS | SMS P/F |
|---|---|---|---|---|---|---|---|---|---|
| 1 | 112 | 113 | 114 | 1 | 0 | 1 | 3 | 0 | 1 |
| 2 | 113 | 114 | 115 | 2 | 1 | 1 | 5 | 2 | 1 |
| 3 | 114 | 115 | 118 | 4 | 2 | 1 | 6 | 1 | 1 |
| 4 | 115 | 118 | 119 | 3 | -1 | 1 | 4 | -2 | 0 |

**Figure 9.** A sample outcome of the communication pattern evaluations on weekdays

|   | weekend day | Outgoing Calls | Delta Calls | Calls P/F | Outgoing SMS | Delta SMS | SMS P/F |
|---|---|---|---|---|---|---|---|
| 1 | 110 | 0 | 0 | 1 | 0 | 0 | 1 |
| 2 | 116 | 1 | 1 | 1 | 7 | 7 | 1 |
| 3 | 117 | 3 | 2 | 1 | 4 | -3 | 1 |
| 4 | 123 | 0 | -3 | 0 | 0 | -4 | 1 |

**Figure 10.** A sample outcome of the communication pattern evaluations on weekends